\documentclass[journal]{IEEEtran}
\usepackage[numbers,sort&compress]{natbib}
\usepackage{amsfonts}
\usepackage[cmex10]{amsmath}
\usepackage{array}
\usepackage{mdwmath}
\usepackage{mdwtab}
\usepackage{eqparbox}
\usepackage[font=normalsize]{caption}
\usepackage{hyperref}
\usepackage{stfloats}
\usepackage{url}
\usepackage{amssymb}
\usepackage{float}
\usepackage{indentfirst}
\usepackage{makeidx}
\usepackage{tabularx}
\usepackage{cases}
\usepackage{multirow}
\usepackage{booktabs}
\usepackage{diagbox}
\usepackage{mathtools}
\usepackage{color}
\usepackage{stfloats}
\usepackage{lipsum}
\usepackage{amsfonts}
\usepackage{bm}
\usepackage{dsfont}
\usepackage{lipsum}
\usepackage{amsthm}
\usepackage{graphicx, epsfig, subfigure}
\usepackage[font=small]{caption}
\allowdisplaybreaks[4]

\makeatletter

\newif\if@restonecol
\makeatother

\usepackage[linesnumbered,ruled,vlined]{algorithm2e}
\usepackage{algpseudocode}
\usepackage{amsmath, bm}

\begin{document}
\title{Compositional Distributed Learning for Multi-View Perception: A Maximal Coding Rate Reduction Perspective}
\author{Zhuojun~Tian and Mehdi~Bennis, \textit{Fellow, IEEE}
	\thanks{Z.~Tian, and M.~Bennis are with the Center for Wireless Communications, University of Oulu, Oulu 90014, Finland. Email: \{zhuojun.tian, mehdi.bennis\}@oulu.fi. This work was supported in part by the ERA-NET CHIST-ERA Project MUSE-COM2 and the Research Council of Finland (former Academy of Finland) Project Vision-Guided Wireless Communication. The code is available on \href{ https://github.com/ZhuoJTian/Compositional-MCR2}{https://github.com/ZhuoJTian/Compositional-MCR2}.}}
\maketitle

\begin{abstract}
In this letter, we formulate a compositional distributed learning framework for multi-view perception by leveraging the maximal coding rate reduction principle combined with subspace basis fusion. 
In the proposed algorithm, each agent conducts a periodic singular value decomposition on its learned subspaces and exchanges truncated basis matrices, based on which the fused subspaces are obtained. By introducing a projection matrix and minimizing the distance between the outputs and its projection, the learned representations are enforced towards the fused subspaces. 
It is proved that the trace on the coding-rate change is bounded and the consistency of basis fusion is guaranteed theoretically. Numerical simulations validate that the proposed algorithm achieves high classification accuracy while maintaining representations' diversity, compared to baselines showing correlated subspaces and coupled representations. 
\end{abstract}
\begin{IEEEkeywords}
Distributed learning, multi-view perception, maximal coding rate reduction, subspace learning.
\end{IEEEkeywords}

\vspace{-0.5cm}
\section{Introduction} \label{sec1} 
In conventional distributed learning, each agent has access to its own full-view training data and cooperates with others to achieve consensus. However, in large scale scenarios, each agent may only observe a partial view of the global environment due to limited sensing capability or various geographical locations \cite{tian2023distributed1, tian2025communication}, leading to the multi-view perception problem \cite{tian2023distributed2}. Through information exchange among agents, compositional distributed learning seeks to integrate partial local knowledge into a global understanding of the environment.

There have been extensive studies on multi-view perception. 
Conventional methods include subspace-based approaches \cite{hotelling1992relations, carroll1968generalization, li2019reciprocal, li2019flexible}, such as canonical correlation analysis (CCA) \cite{hotelling1992relations} and its generalized version GCCA \cite{carroll1968generalization}, as well as spectral-based methods \cite{kumar2011co, xia2010multiview}. 
With the development of deep learning, high-level associations among multi-view data can be better captured through non-linear neural networks \cite{andrew2013deep, benton2017deep, shaham2018spectralnet, guo2022multi, lou2025parameter, guo2025globality}. 
Deep CCA \cite{andrew2013deep} and DGCCA \cite{benton2017deep} adopt a common strategy of learning joint representations across multiple views at a higher level, while capturing view-specific features in the lower layers. 
Another line of research leverages auto-encoders \cite{wu2018multimodal, lee2020private, LawryAguila2023} to construct a shared latent space from multi-view inputs. Although insightful, they are primarily designed for centralized settings and overlook data privacy concerns in distributed environments. To address multi-view datasets collected by distributed agents, recent studies have introduced federated multi-view clustering (FedMVC) \cite{huang2020federated, huang2022efficient, chen2023federated, chen2024bridging}. The authors in \cite{chen2023federated} leverage global self-supervised information to extract complementary cluster information, while the method in \cite{chen2024bridging} coined as FMCSC further considers hybrid views using contrastive learning techniques. These methods however fail to exploit the structure of representation spaces and lack interpretability.

The authors in \cite{yu2020learning} introduced the Maximal Coding Rate Reduction (MCR$^2$) principle generating independent feature subspaces where features are distributed isotropically. As a result, the principal directions within these subspaces become more stable and uniformly distributed. 
Inspired by the stability and interpretability of the captured subspaces, our work introduces MCR$^2$ as a discriminative criterion for multi-view feature fusion, in order to provide a rigorous information-theoretic interpretation. 
Given the well-structured principle components of the learned subspaces, we design a periodic basis fusion procedure to compose the local subspaces into global one.
Our contributions can be summarized as follows:
\begin{itemize}
    \item{We formulate a distributed multi-view perception problem leveraging the MCR$^2$ principle. By utilizing the isotropical properties of the subspaces, we design a periodic basis fusion to integrate the local subspaces, and the projection loss to adjust the output features' subspace.}
    \item{The bound on the variation of the coding rate is characterized by the projection residual energy. We further establish the convergence rate of the fused subspace matches that of the local covariance estimation error.}
    \item{We evaluate the algorithm on multi-view perception tasks and benchmark it against several baselines, demonstrating that the output representations preserve the diversity and discriminability properties of the MCR$^2$ principle.}
\end{itemize}

\vspace{-0.5cm}
\section{System Model and Problem Formulation}\label{Sec2}
\subsection{System Model}
Consider a decentralized multi-agent communication network, which can be represented by an undirected graph $\mathcal{G}$ with $N$ distributed agents/nodes.
Each agent has access to a partial view of the global objects, collecting local dataset denoted by $\mathcal{D}_i=\{\mathcal{X}_i, \mathcal{Y}_i\}$.
The output of the representation learning neural network in agent $i$ is defined as $\bm{Z}_i\in\mathbb{R}^{d\times m_i}$, where $d$ is the dimension of the output feature and $m_i$ is the number of data samples in node $i$. 
If we denote the representation learning neural network (encoder) in agent $i$ by $f_i$, which is parameterized by $\bm{\theta}_i$, then the output can be represented as $\bm{Z}_i=f_i(\bm{X}_i, \bm{\theta}_i)$.

\subsection{Maximal Coding Rate Reduction}
The compactness of the learned features $\bm{Z}$ as a whole can be measured by the average coding length per sample when the sample size is large enough, i.e., the coding rate subject to the distortion \cite{ma2007segmentation, yu2020learning}, is given by:
\begin{equation}
    \label{eqR}
    R(\bm{Z}, \epsilon) = \frac{1}{2}\log\det(\bm{I}+\frac{d}{m\epsilon^2}\bm{Z}\bm{Z}^T),
\end{equation}
{which represents the minimal number of binary bits needed to encode $\bm{Z}$ such that the expected decoding error is less than $\epsilon$ \cite{ma2007segmentation}.}
Considering that the generated features $\bm{Z}$ have multiple classes from different subspaces, w.r.t. this partition, the average number of bits per sample (the coding rate) is given in (\ref{eqRc}), where $\bm{\Pi}_k$ is a diagonal matrix whose diagonal entries indicate the membership of the samples in the multiple classes. In this regard, the label serves as side information.
\begin{equation}
    \label{eqRc}
    R^c(\bm{Z}, \epsilon|\bm{\Pi})\!=\!\sum_{k=1}^K\frac{tr(\bm{\Pi}_k)}{2m}\log\det(\bm{I}+\frac{d}{tr(\bm{\Pi}_k)\epsilon^2}\bm{Z}\bm{\Pi}_k\bm{Z}^T).
\end{equation}
To maximize the discrimination of the features among different classes, the whole space of $\bm{Z}$ must be as large as possible, {so that features of different samples are maximally incoherent to each other}. On the other hand, within each class, the subspace should be of small volume to make the representation compact and more correlated. Therefore, the loss function on the principle of Maximal Coding Rate Reduction can be expressed as follows \cite{yu2020learning}:
\begin{equation}
    \label{MCR2}
    \max\quad\mathcal{M}(\bm{Z})=R(\bm{Z}, \epsilon) - R^c(\bm{Z}, \epsilon|\bm{\Pi}).
\end{equation}
MCR$^2$ principle produces features spaces that are between-class discriminative and maximally diverse within each class \cite{yu2020learning}. Each subspace corresponds to a class, where the principal directions are more stable and evenly distributed. On this basis, we develop the algorithm in the following section.

\vspace{-0.2cm}
\section{Proposed Algorithm}\label{algo}
\begin{figure}[t]
    \centering
    \includegraphics[width=0.45\textwidth]{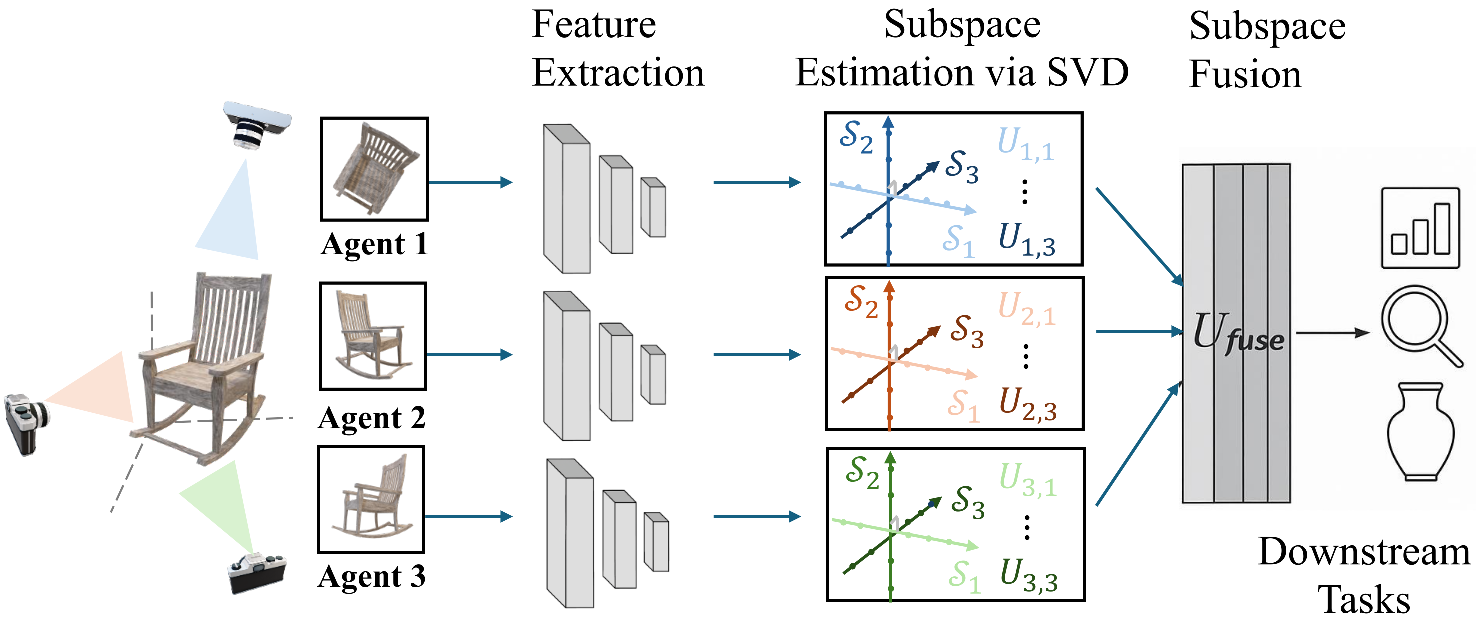}
    \caption{Illustration of the proposed multi-view perception framework.}
    \label{fig1}
    \vspace{-0.5cm}
\end{figure}

In this section, we develop the compositional distributed learning framwework,
as shown in Fig. \ref{fig1}.
In each agent, the encoder learns well-structured subspaces through the MCR$^2$ principle, which are composed through the basis fusion and the designed projection loss term.

\vspace{-0.2cm}
\subsection{Periodic basis fusion}
Specifically, the output features of agent $i$ corresponding to the $k$-th class are denoted by $\bm{Z}_{i,k}$. 
This formulated subspace can be represented by its principal components, which can be obtained through the singular value decomposition (SVD), i.e., $\bm{Z}_{i,k} = \bm{U}_{i,k}\bm{\Sigma}_{i,k}\bm{V}_{i,k}^T$.
Here $\bm{U}_{i,k}\in\mathbb{R}^{d\times d}$ is the left singular vectors with orthonormal basis for the column space giving the feature directions. The principle components of the corresponding subspace can be obtained through the leading $p_{k}$ columns of $\bm{U}_{i,k}$, denoted by $\hat{\bm{U}}_{i,k}$. The principle components represent the corresponding subspace. Sorting the singular values in $\bm{\Sigma}_{i,k}$ in descending order yeilds $\hat{\bm{U}}_{i,k}=\bm{U}_{i,k}[:,0:p_{k}]$.

After conducting SVD on local feature spaces, each agent transmits its principle components to the central server or other nodes for basis fusion.
In the Federated Learning framework, the central server fuses the received basis from all agents, while in decentralized settings, each agent conducts fusion locally with the received information, using for instance multi-hop transmission.
These basis from all agents can be concatenated as $\tilde{\bm{U}}_{k}=[\hat{\bm{U}}_{1,k}, ..., \hat{\bm{U}}_{N,k}]$  for each class $k$.
To obtain the fused subspace, the server or each agent needs to take another SVD on the concatenated basis matrix, i,e, $\tilde{\bm{U}}_{i} = \bar{\bm{U}}_{i}\bar{\bm{\Sigma}}_{i}\bar{\bm{V}}_{i}^T$.
Through importance ranking and selecting the first $P_{k}$ columns of $\bar{\bm{U}}_{i}$, the fused basis for the composed subspace can be obtained through:
\vspace{-0.2cm}
\begin{equation}
    \label{get_Ufuse}
    \hat{\bm{U}}_{fuse,k}=\bar{\bm{U}}_{k}[:,0:P_{k}].
    \vspace{-0.2cm}
\end{equation}
Such SVD operation ensures the fused basis matrix with global orthogonalization, where redundancy can be removed while keeping the complementary information. 
{
In each round, the additional overall computational cost resulted from these truncated SVD operations can be approximated by $\mathcal{O}(\sum_{k}(Mdp_k+Ndp_kP_k))$, with $M=\sum_i m_i$.
}

\vspace{-0.2cm}
\subsection{Loss function design}
Note that the basis fusion is an external and non-differentiable subspace estimation step, which cannot be directly involved in updating the outputs features.
To solve this issue and update the output subspace, we design a projection loss term, which ensures the output features are close to the fused subspaces.

Specifically, for the $k$-th class in agent $i$, we define $\bm{P}_{k}=\bm{U}_{fuse,k}\bm{U}_{fuse,k}^T$. Then Lemma \ref{l1} can be obtained.
\newtheorem{lemma}{Lemma}
\newtheorem{theorem}{Theorem}
\newtheorem{corollary}{Corollary}
\begin{lemma}\label{l1}
$\bm{P}_{k}\in\mathbb{R}^{d\times d}$ is the orthogonal projection operator satisfying $\bm{P}_{k}^2=\bm{P}_{k}$ and $\bm{P}_{k}^T=\bm{P}_{k}$ and projects the vectors to the subspace formulated by the basis $\bm{U}_{fuse,k}$.
\end{lemma}
Given $\bm{P}_{k}$, the output features in agent $i$ corresponding to class $k$, denoted by $\bm{Z}_{i,k}$, can be projected to the fused subspace through $\bm{P}_{k}\bm{Z}_{i,k}$. 
Then to make the learned features close to the fused subspace, we design the following projection loss term by minimizing the $\ell_2$ distance, i.e., $\sum_{k=1}^K\|\bm{Z}_{i,k}-\bm{P}_{k}\bm{Z}_{i,k}\|_F^2$.
Adding the projection loss term to (\ref{MCR2}), the local loss function of agent $i$ can be formulated as: 
\begin{align}
    \label{local_loss}
    & \min_{\bm{Z}_i} R^c(\bm{Z}_i, \epsilon|\bm{\Pi}_i) - R(\bm{Z}_i, \epsilon) + \lambda\sum_{k=1}^K\|\bm{Z}_{i,k}-\bm{P}_{k}\bm{Z}_{i,k}\|_F^2, \notag\\
    &\text{s.t.}\quad \|\bm{Z}_{i,k}\|_F^2 = m_{i,k}, \forall 1\le k\le K, 
\end{align}
where $\lambda$ is the parameter controlling the influence of the projection term. In (\ref{local_loss}), the last regularization term measures the distance to the fused subspace, leading to subspace alignment and discriminative learning together with the MCR$^2$ principle.
The constraint seeks to ensure the reduction is comparable across different representations. This can be achieved by normalizing each feature to lie on the unit sphere \cite{yu2020learning}, which can be implemented by adding a normalization function to the output layer.

Based on the basis fusion and the designed loss function, the algorithm can be summarized in Algorithm \ref{Alg1}.
\begin{algorithm}
	\caption{\textbf{Compositional Distributed Learning for Multi-View Perception (CDL-MVP)}}\label{Alg1}
	\For{node $i=1,2,\dots, N$ in parallel}
	{
		\textbf{Initialize} the local parameters of encoder $\bm{\theta}_i$, the dimension of the output feature $d$, $p_{i,k}$ and $P_{i,k}$. $t=0$ for all classes. \\
            \textbf{Take} SVD on the output features in each class, select the first $p_{k}$ columns to get $\hat{\bm{U}}_{i,k}$, and \textbf{transmit} $\hat{\bm{U}}_{i,k}$ to all other nodes.\\
	}
	\While{not converge}
	{
		$t=t+1$\\
		\For{node $i=1,2,\dots, N$ in parallel}
		{
			\For{class $k=1,2,\dots, K$ in parallel}
                {
                    \textbf{Fuse} the received basis, get $\hat{\bm{U}}_{fuse,k}^{(t)}$ according to (\ref{get_Ufuse}) and compute the projection matrix $\bm{P}_{k}^{(t)}$.\\
                }
                
                \For{inner step $t'=1,\dots, T'$}
			{
			    \textbf{Update} $\bm{\theta}_i$ with stochastic gradient descent based on the loss in (\ref{local_loss}).\\
			}
            
			\textbf{Obtain} $\bm{Z}_{i,k}^{(t)}\in\mathbb{R}^{d\times m_{i,k}}$ for all classes with all of the training data samples.\\
                \textbf{Take} SVD on the output features in each class, select the first $p_{k}$ columns to get $\hat{\bm{U}}_{i,k}^{(t)}$, and \textbf{transmit} $\hat{\bm{U}}_{i,k}^{(t)}$ to all other nodes.\\
		}
	}
	\textbf{Output} the trained local encoders and the resultant representations $\bm{Z}_i$ for all agents.
\end{algorithm}
\vspace{-1.0cm}

\section{Theoretical Analysis}
In this section, we provide a theoretical analysis of the proposed algorithm.
Due to space limitations, the full proof is provided in the supplementary material.
Define the local projection matrix as the diagonal matrix for all $\bm{P}_{i,k}$, i.e., $\tilde{\bm{P}}_i=\textbf{diag}(\{\bm{P}_{i,k}\})$.
The projected features in one agent can be thus denoted by \(\bm{Z}^P_i := \tilde{\bm{P}}_i\bm{Z}_i\).
For node $i$, define the projection residual energy of the features over all classes and within each class respectively as
\[
\varepsilon_i \;=\; \|(\bm{I}-\tilde{\bm{P}}_i)\bm{Z}_i\|_F^2, \qquad
\varepsilon_{i,k} \;=\; \|(\bm{I}-\bm{P}_{i,k})\bm{Z}_{i,k}\|_F^2.
\]

\begin{theorem}[Linear trace bound on coding-rate change]
\label{t1}
Given the above definition, the changes in the MCR$^2$ loss due to projection is tightly bounded by the projection residual energy:
\[
\big|\mathcal{M}(\bm{Z}_i) - \mathcal{M}(\bm{Z}_i^P)\big|
\le \frac{d}{m_i\epsilon^2}\,\varepsilon_i \;+\; \sum_{k=1}^K \frac{d}{m_{i}\epsilon^2}\,\varepsilon_{i,k}.
\]
Suppose that during training we reach a point where $\varepsilon_i \le \delta_i$ and $\varepsilon_{i,k} \le \delta_{i,k}$, where \(\delta_i,\{\delta_{i,k}\}\) are small. Then in each agent, the MCR\(^2\)-difference is \(O(\delta+\sum_k\delta_k)\).
\end{theorem}

The bound in Theorem~\ref{t1} offers a direct certificate on how accurately the projected-space MCR\(^2\) approximates the true MCR\(^2\), through monitoring the projection residual energy during training. 
Moreover, Theorem~\ref{t1} builds a simple yet interpretable connection between the reconstruction penalty and the fidelity of evaluating MCR\(^2\) on the fused subspace. 

Before Theorem \ref{t2}, we first give some definitions for better illustration.
Let $\mathcal{S}^* \subset \mathbb{R}^d$ denote the true global discriminative subspace with dimension $\dim(\mathcal{S}^*) = R$, composed of $K$ orthogonal subspaces corresponding to $K$ classes \cite{yu2020learning}. Let $\bm{U}^* \in \mathbb{R}^{d \times R}$ be an orthonormal basis of $S^*$. For each agent $i \in \{1,\dots,N\}$, let $\bm{U}_i^* \in \mathbb{R}^{d \times r_i}, r_i \geq 1$ be the population-optimal local subspace obtained by solving the local MCR$^2$-type optimization. 
$\operatorname{range}(\bm{U})$ denotes the column space of a matrix $\bm{U}$, and we assume $\operatorname{range}(\bm{U}_i^*) \subseteq \mathcal{S}^*$, i.e., there exists a column-orthogonal matrix $\bm{O}_i \in \mathbb{R}^{R \times r_i}$ such that $\bm{U}_i^* = \bm{U}^* \bm{O}_i$.
Define the \emph{coverage matrix}
\[
  \bm{M} := [\, \bm{O}_1, \bm{O}_2, \dots, \bm{O}_N \,] \in \mathbb{R}^{R \times r_{\rm tot}},
  \qquad r_{\rm tot} := \sum_{i=1}^N r_i.
\]
At the sample level, each agent computes an estimated subspace with basis matrix $\widehat{\bm{U}}_i \in \mathbb{R}^{d \times r_i}$. Denote the estimation error of the corresponding covariance-type matrices by $\Delta_i := \|\widehat{\bm{\Sigma}}_i - \bm{\Sigma}_i\|$. 
Define the ideal (population) concatenation and the sample concatenation as:
\[
  \bm{B}^* := [\, \bm{U}_1^*, \dots, \bm{U}_N^* \,] = \bm{U}^* \bm{M}, 
  \qquad
  \bm{B} := [\, \widehat{\bm{U}}_1, \dots, \widehat{\bm{U}}_N \,].
\]
Let $\bm{B} = \bar{\bm{U}} \bm{\Sigma}\bm{V}^\top$ be the singular value decomposition. We define the fused subspace estimate by $\widehat{\bm{U}}_{\mathrm{fuse}} := \bar{\bm{U}}_{[:,1:R]}$.
Define 
\[
\sin\Theta(\widehat{\bm{U}}_i, \bm{U}_i^*) := \operatorname{diag}(\sin\theta_1, \dots, \sin\theta_{r_i}),
\]
where $\theta_1, \dots, \theta_{r_i}$ are the principal angles between the two subspaces of the same dimension \cite{davis1970rotation}. Finally, recall the Grassmann distance between two subspaces $\mathcal{S}, \mathcal{T}$ \cite{hamm2008grassmann}:

\[
  d_{\rm Gr}(\mathcal S,\mathcal T) := \|\sin\Theta(\mathcal S,\mathcal T)\|_2
  = \| P_\mathcal{S} - P_\mathcal{T} \|_2 .
\]

\begin{theorem}[Consistency of SVD Fusion]\label{t2}
Denote the $R$-th largest singular value of $\bm{M}$ by $\sigma_R(\bm{M})$. Assume $\sigma_R(\bm{M}) \geq \beta > 0$, i.e., the local subspaces collectively span $\mathcal{S}^*$ with non-degenerate coverage. Additionally, there exists a constant $L < \infty$ and a fixed eigengap $\mathrm{gap} > 0$
such that
\[
  \|\sin\Theta(\widehat{\bm{U}}_i, \bm{U}_i^*)\| \;\leq\; L \, \Delta_i, 
  \qquad \forall i=1,\dots,N.
\]
Under the coverage and spectral stability assumptions above, there exists a
constant $C>0$ depending only on $(L,\beta,N)$ such that
\[
  d_{\rm Gr}\!\left(\operatorname{range}(\widehat{\bm{U}}_{\rm fuse}),\, \mathcal{S}^*\right)
  \;\leq\; C \cdot \max_{1\leq i\leq N} \Delta_i .
\]
In particular, if $\Delta_i = o_P(1)$ for all agents, then $d_{\rm Gr}\!\left(\operatorname{range}(\widehat{\bm{U}}_{\rm fuse}),\, \mathcal{S}^*\right)
  = o_P(1)$.
\end{theorem}
Theorem \ref{t2} provides a rigorous justification of the SVD fusion procedure. 
We show that under mild assumptions, the fused subspace estimate obtained from the local agents converges to the true global discriminative subspace at the
same rate as the local covariance estimation error. 

\section{Numerical Experiments}\label{experiment}
\setcounter{figure}{2}
\begin{figure*}[!htp]
\centering
\subfigure[CDL-MVP]{
\begin{minipage}[t]{0.13\textwidth}
\centering
\includegraphics[width=\textwidth]{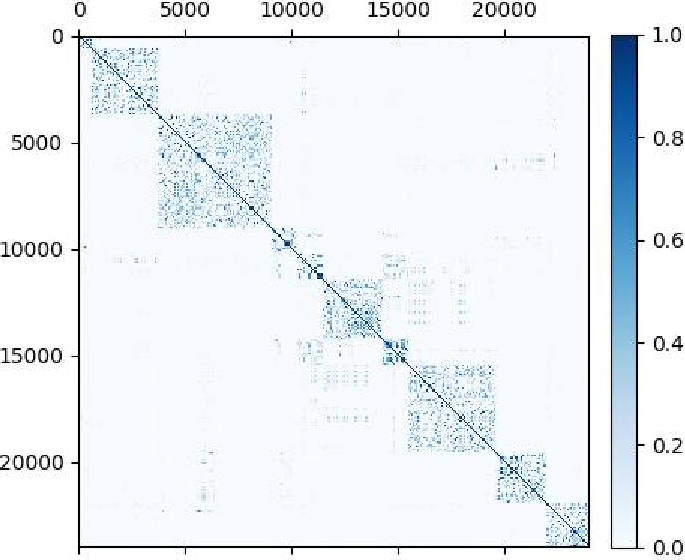}
\end{minipage}%
}%
\subfigure[IndepMCR]{
\begin{minipage}[t]{0.13\textwidth}
\centering
\includegraphics[width=\textwidth]{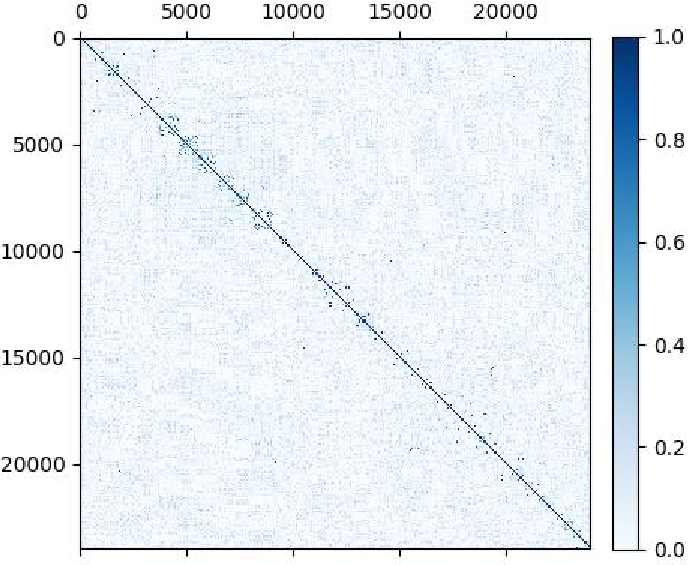}
\end{minipage}%
}%
\subfigure[CE-SVD]{
\begin{minipage}[t]{0.13\textwidth}
\centering
\includegraphics[width=\textwidth]{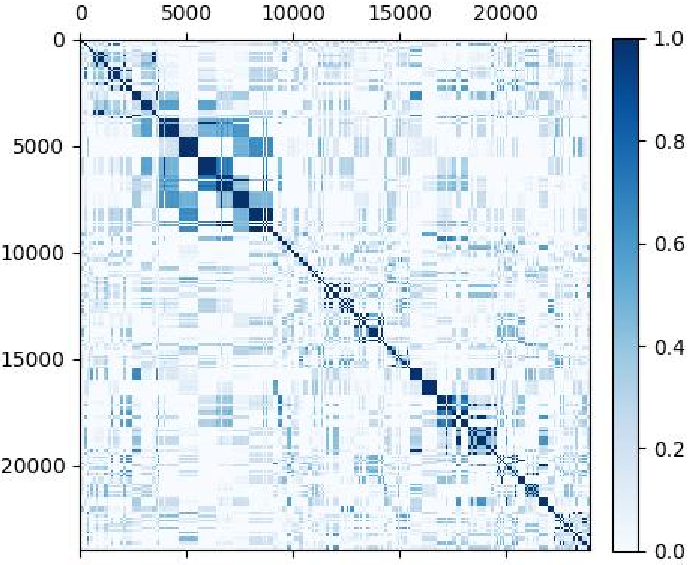}
\end{minipage}
}%
\subfigure[CE-Avg]{
\begin{minipage}[t]{0.13\textwidth}
\centering
\includegraphics[width=\textwidth]{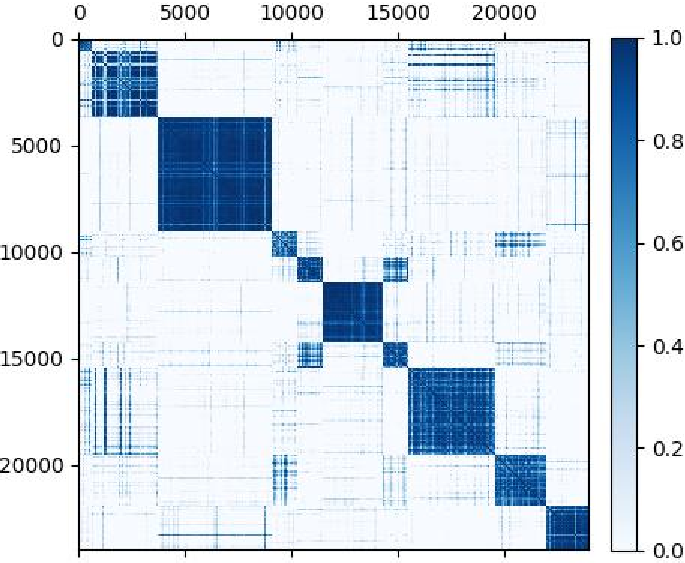}
\end{minipage}
}%
\subfigure[DGCCA]{
\begin{minipage}[t]{0.13\textwidth}
\centering
\includegraphics[width=\textwidth]{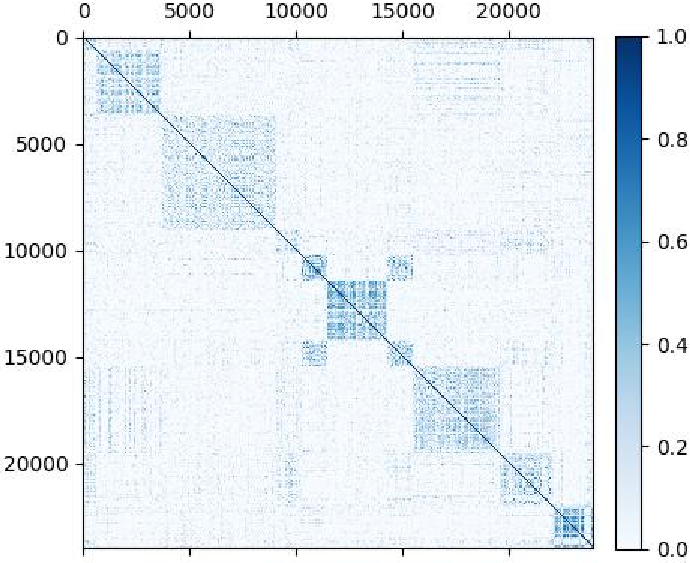}
\end{minipage}
}%
\subfigure[MVAE]{
\begin{minipage}[t]{0.13\textwidth}
\centering
\includegraphics[width=\textwidth]{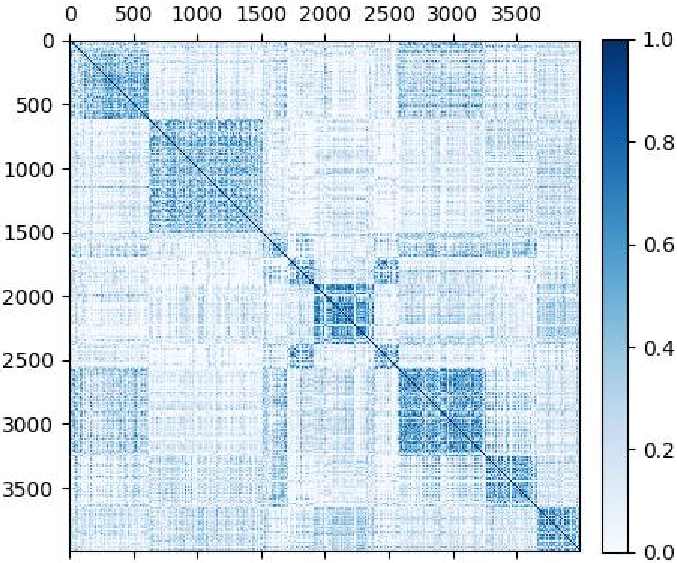}
\end{minipage}
}%
\subfigure[FMCSC]{
\begin{minipage}[t]{0.13\textwidth}
\centering
\includegraphics[width=\textwidth]{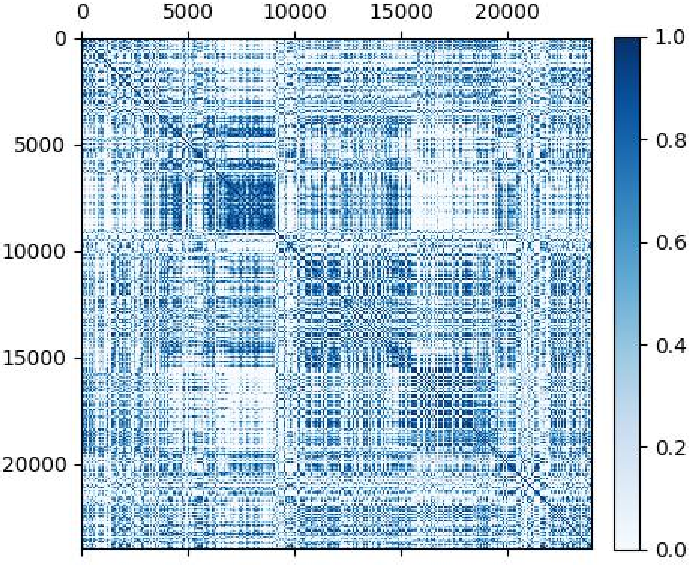}
\end{minipage}
}%
\vspace{-0.2cm}
\centering
\caption{Cosine similarity of the learned representations for ModelNet-10.}\label{fig_exp2}
\vspace{-0.5cm}
\end{figure*}

In this section, we evaluate the performance of the proposed algorithm in the multi-view perception scenario, including both 2 dimensional (2D) images and 3 dimensional (3D) objects, {with one A100 Tesla GPU}. 
Specifically, in the 2D scenario, we consider CIFAR-10 dataset, where $4$ agents have access to different regions ($18\times18$) of the images. The agents have different neural network (NN) architectures: ResNet18, ResNet34, VGG11 and VGG16.
In the 3D scenario, we use the ModelNet-10 dataset, where $6$ agents take images of the 3D objects from different views. The agents share the same NN model, consisting of $4$ convolutional layers, followed by one flattened layer and one linear layer. The output dimension of the features is set to $d=64$. Through experiments, we set $p_{i,k}=10$, $P_{i,k}=16$ for all agents $i$ and all classes $k$.

We compare the proposed algorithm with independent MCR$^2$ (IndepMCR) and cross-entropy loss with SVD and basis fusion (CE-SVD).
For ModelNet-10 where agents share the same model architecture, we additionally compare with cross-entropy with averaging (CE-Avg), as well as other state-of-the-art algorithms, including centralized DGCCA \cite{benton2017deep}, MVAE \cite{wu2018multimodal, LawryAguila2023} and distributed FMCSC \cite{chen2024bridging}.
For the proposed algorithm CDL-MVP, we set the initial learning rate as $0.01$ for CIFAR-10 dataset and $0.001$ for ModelNet-10, with $10^{-5}$ weight decay, and use the Adam optimizer. For both datasets, in the first $4000$ epochs, $\lambda$ is set to $1.0$, while in the last $2000$ epochs, $\lambda$ is set to $100.0$. {The batch size is set to $128$ and the agents exchange their information after each local epoch.}
The cosine-similarity results for CIFAR-10 dataset are shown in Fig. \ref{fig_exp1} and those for ModelNet-10 are shown in Fig. \ref{fig_exp2}.
\setcounter{figure}{1}
\begin{figure}[!htbp]
\vspace{-0.3cm}
\centering
\subfigure[CDL-MVP]{
\begin{minipage}[t]{0.33\linewidth}
\centering
\includegraphics[width=\textwidth]{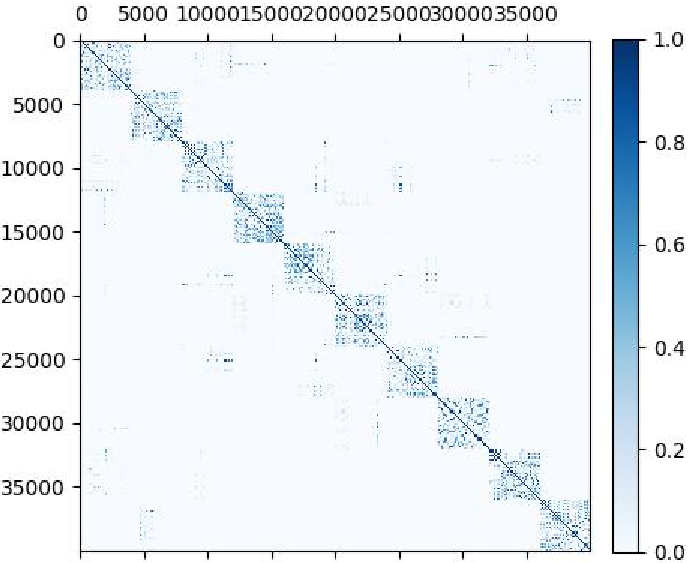}
\end{minipage}%
}%
\subfigure[IndepMCR]{
\begin{minipage}[t]{0.33\linewidth}
\centering
\includegraphics[width=\textwidth]{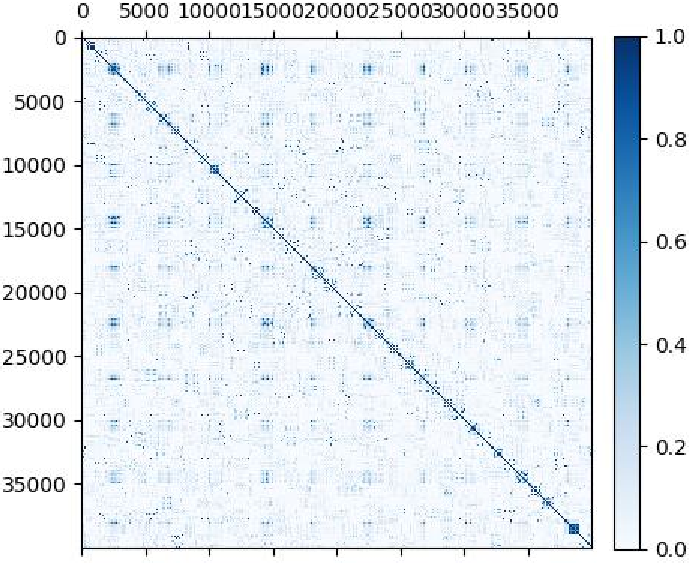}
\end{minipage}%
}%
\subfigure[CE-SVD]{
\begin{minipage}[t]{0.33\linewidth}
\centering
\includegraphics[width=\textwidth]{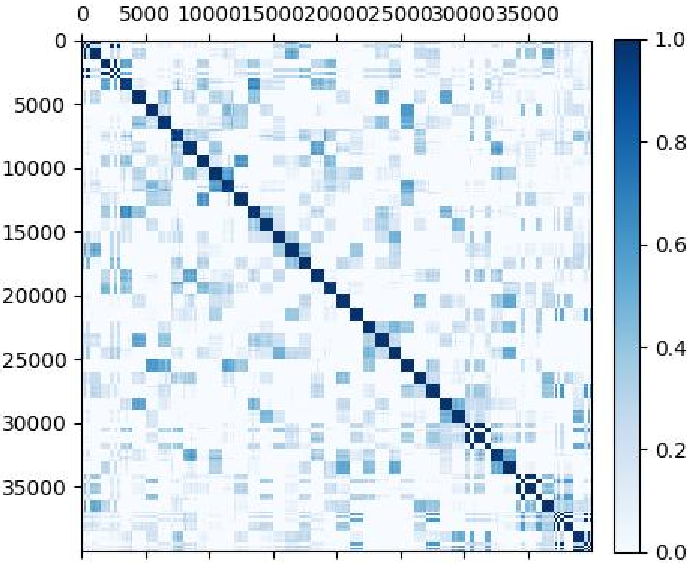}
\end{minipage}
}%
\vspace{-0.2cm}
\centering
\caption{Cosine similarity of the learned representations for CIFAR-10.}\label{fig_exp1}
\vspace{-0.3cm}
\end{figure}

The cosine similarity results of the proposed CDL-MVP shown in Fig. \ref{fig_exp1}(a) and \ref{fig_exp2}(a) illustrate that the fused representations align with the diverse and discriminative properties of MCR$^2$ principle. 
Comparing the results between Fig. \ref{fig_exp1} and \ref{fig_exp2}, we see that the cooperation among agents through the basis fusion and designed projection loss is effective, where the subspaces within each agent are composed into global subspaces. 
The results in Fig. \ref{fig_exp1}(c) and \ref{fig_exp2}(c) validate that the subspaces learned by MCR$^2$ principle can be fused through the principle components while the cross-entropy collapses the feature spaces, disallowing composition. The collapse of spaces with cross entropy is also shown in Fig. \ref{fig_exp2}(d), where the within-class features converge to their respective class means. Thus, cross-entropy can only deal with the classification task compromising the intrinsic structure of individual data samples. In Fig. \ref{fig_exp2}(e) and (f), the results of the centralized DGCCA and MVAE exhibit similar performance as CDL-MVP. However, the learned subspaces among different classes are still correlated compared with those in Fig. \ref{fig_exp2}(a). 

For the ModelNet-10 dataset, Table \ref{table_1} quantitatively shows the performance on testing dataset: Acc is the classification accuracy; SIS is the cosine similarity among different views of the same objects; DIS is the cosine similarity among the different objects within one class; FR is the Fisher ratio, defined as the ratio of between-class variance to within-class variance. The results show that the proposed CDL-MVP can achieve comparable accuracy while maintaining the diversity of representations within each class (as indicated by DIS) and among the different views of the same object (as shown by SIS). 
{Both CDL-MVP and IndepMCR exhibit relatively low values of SIS, DIS, and FR, which can be attributed to the first term (\ref{eqR}) expanding the overall feature space and preserving sample diversity within each subspace. To further enhance the correlation among representations of the same image, it may be beneficial to incorporate an additional contrastive loss term \cite{lou2025parameter}.}
Here, MVAE composes the outputs of different views into one common feature, thus has no SIS and FR value. 
\begin{table}[!htp]
\centering
\caption{Comparison on different measurements.}\label{table_1}
\small
\begin{tabular}{lllll}
\hline
  {}       & Acc  & SIS & DIS & FR \\ \hline
 CDL-MVP   & 0.8533 & 0.0043 & \textbf{0.0240} & \textbf{0.8851}  \\
 IndepMCR & 0.7094 & \textbf{0.0016} & 0.0256 & 2.2711      \\
CE-SVD   & 0.7535 & 0.0516 & 0.1780 & 2.1147\\
 CE-Avg & 0.8592 & \textbf{0.8127} & 0.7484 & 1.5092     \\
 DGCCA & 0.7819 & 0.4788 & 0.2042 & 1.4973  \\
 MVAE & \textbf{0.8733} & --- & 0.4395 & ---     \\
 FMCSC & 0.6428 & 0.3161 & 0.3307 & 1.4320     \\\hline
\end{tabular}
\vspace{-0.5cm}
\end{table}

\section{Conclusion}\label{conclusion}
In this letter, we present a compositional distributed algorithm for multi-view perception, that leverages the structural subspaces derived from the MCR$^2$ principle. 
We introduce periodic basis fusion alongside a tailored projection loss, enabling the integration of local subspaces into global representations. The proposed CD-MVP framework is validated both theoretically and empirically through comparisons with existing methods. Future directions include {deriving the results for gossip-based decentralized learning}, optimizing the computational efficiency of the basis fusion process and adapting the algorithm for generative applications.

\newpage
\section*{Supplementary Materials}
\begin{lemma}
\label{l2}
For any orthogonal projection, under the definition in (\ref{eqR}), we have $R(\bm{Z}^P_i)\le R(\bm{Z}_i)$.
\end{lemma}

\subsection{Proof of Lemma \ref{l2}}
\begin{proof}
Let \(S=\tfrac{\alpha}{n}ZZ^\top\succeq 0\) and \(S_P=\tfrac{\alpha}{n}PZZ^\top P = \tfrac{\alpha}{n}P S P \succeq 0\).
Because \(P S P \preceq S\) in the Loewner order, the matrix function \(A\mapsto\log\det(I+A)\) is monotone increasing on the PSD cone, hence
\[
\log\det(I+S_P) \le \log\det(I+S).
\]
Multiplying by \(1/2\) yields \(R(Z_P)\le R(Z)\). The same argument applied to each class-block \(Z_y\) gives the second claim.
\end{proof}

\subsection{Proof of Theorem \ref{t1}}
\begin{proof}
Let \(\bm{A}=\tfrac{d}{m_i\epsilon^2}\bm{Z}_i\bm{Z}_i^\top\) and \(\bm{B}=\tfrac{d}{m_i\epsilon^2}\tilde{\bm{P}}_i\bm{Z}_i\bm{Z}_i^\top \tilde{\bm{P}}_i\). Both are positive semi-definite and \(\bm{B}\preceq \bm{A}\). Using the scalar inequality \(\log(1+t)\le t\) for \(t>-1\), applied to eigenvalues, we obtain
\begin{align}
&\log\det(\bm{I}+\bm{A})-\log\det(\bm{I}+\bm{B})\\
&= \sum_{i}\big(\log(1+\lambda_i(\bm{A})) - \log(1+\lambda_i(\bm{B}))\big)\notag\\
&\le \sum_{i}\big(\lambda_i(\bm{A}) - \lambda_i(\bm{B})\big)= \operatorname{Tr}(\bm{A}-\bm{B}).\notag
\end{align}
Dividing both sides by \(2\) gives
\[
R(\bm{Z}_i)-R(\bm{Z}_i^P) \le \tfrac{1}{2}\operatorname{Tr}(\bm{A}-\bm{B}).
\]
Additionally, we have
\begin{align}
&\operatorname{Tr}(\bm{A}-\bm{B}) = \frac{d}{m_i\epsilon^2}\operatorname{Tr}\big(\bm{Z}_i\bm{Z}_i^\top - \tilde{\bm{P}}_i\bm{Z}_i\bm{Z}_i^\top \tilde{\bm{P}}_i\big)\notag\\
&= \frac{d}{m_i\epsilon^2}\operatorname{Tr}\big((\bm{I}-\tilde{\bm{P}}_i)\bm{Z}_i\bm{Z}_i^\top\big)
= \frac{d}{m_i\epsilon^2}\|(\bm{I}-\tilde{\bm{P}}_i)\bm{Z}_i\|_F^2.
\end{align}
Given Lemma \ref{l2}, we have $R(\bm{Z}_i)-R(\bm{Z}_i^P)\ge0$
The per-class bounds follow by replacing \(\bm{Z}_i\) with \(\bm{Z}_{i,k}\). Summing and taking absolute value gives the bounded results.

Immediate from previous results by substituting \(\varepsilon\le\delta\) and the per-class bounds \(\varepsilon_y\le \delta_y\), we can obtain the other results.
\end{proof}

\subsection{Proof of Theorem \ref{t2}}
\begin{proof}
By construction $\bm{B}^* = \bm{U}^* \bm{M}$, we have
\[
  \sigma_R(\bm{B}^*) = \sigma_R(\bm{M}) \;\geq\; \beta > 0, 
  \qquad \sigma_{R+1}(\bm{B}^*) = 0 .
\]
Hence the top-$R$ left singular vectors of $\bm{B}^*$ span exactly $\mathcal{S}^*$, and the singular value gap between the $R$-th and $(R+1)$-th singular values is $\mathrm{gap}_\star = \beta$.

Define $\bm{E} := \bm{B} - \bm{B}^*$. Then
\[
  \|\bm{E}\| 
  \;\leq\; \sqrt{\sum_{i=1}^N \|\widehat{\bm{U}}_i - \bm{U}_i^*\|^2}
  \;\leq\; \sqrt{N}\cdot \max_i \|\widehat{\bm{U}}_i - \bm{U}_i^*\|.
\]
By orthogonal Procrustes \cite{stewart1990matrix} and the spectral stability assumption,
\[
  \|\widehat{\bm{U}}_i - \bm{U}_i^*\| 
  \;\leq\; \sqrt{2}\, \|\sin\Theta(\widehat{\bm{U}}_i, \bm{U}_i^*)\|
  \;\leq\; \sqrt{2}\, L\, \Delta_i.
\]
Therefore we have:
\[
  \|\bm{E}\| \;\leq\; \sqrt{2N}\,L \cdot \max_i \Delta_i.
\]

Wedin's perturbation theorem for singular vectors states that \cite{yu2015useful}
\[
  \|\sin\Theta(\widehat{\mathcal U}, \mathcal U^*)\|
  \;\leq\; \frac{\|\bm{E}\|}{\mathrm{gap}_\star},
\]
where $\widehat{\mathcal U}=\operatorname{range}(\widehat{\bm{U}}_{\rm fuse})$ and
$\mathcal U^*=\operatorname{range}({\bm{U}}^*)=\mathcal{S}^*$. Substituting the bounds above yields
\[
  d_{\rm Gr}\!\left(\operatorname{range}(\widehat{\bm{U}}_{\rm fuse}), \mathcal{S}^*\right)
  \;\leq\; \frac{\sqrt{2N}\,L}{\beta}\cdot \max_i \Delta_i.
\]
This proves Theorem \ref{t2} with $C=\tfrac{\sqrt{2N}\,L}{\beta}$.
\end{proof}

\subsubsection*{Discussion}

\begin{itemize}
\item The coverage constant $\beta$ ensures that the ideal concatenation
$B^*$ has rank $R$ with non-degenerate singular values. Without this,
fusion cannot recover $S^*$.
\item The local stability constant $L$ arises from Davis–Kahan type bounds
for (generalized) eigenspaces, where $L \propto 1/\mathrm{gap}$.
\end{itemize}

\end{document}